\documentclass[aps,prl,twocolumn,superscriptaddress]{revtex4}
\usepackage{graphicx,epsf}

\begin{document}

\title{Land\'{e} g-tensor in semiconductor nanostructures}

\author{T. P. Mayer-Alegre}
\affiliation{Laborat\'orio Nacional de Luz S\'{\i}ncrotron, Caixa Postal 6192 - CEP 13084-971, Campinas, SP , Brazil}
\affiliation{Instituto de F\'isica Gleb Wataghin Universidade Estadual de Campinas, Campinas, SP , Brazil}

\author{F. G. G. Hern\'andez}
\affiliation{Laborat\'orio Nacional de Luz S\'{\i}ncrotron, Caixa
Postal 6192 - CEP 13084-971, Campinas, SP , Brazil}
\affiliation{Instituto de F\'isica Gleb Wataghin Universidade
Estadual de Campinas, Campinas, SP , Brazil}

\author{A. L. C. Pereira}
\affiliation{Laborat\'orio Nacional de Luz S\'{\i}ncrotron, Caixa
Postal 6192 - CEP 13084-971, Campinas, SP , Brazil}
\author{G. Medeiros-Ribeiro}
\affiliation{Laborat\'orio Nacional de Luz S\'{\i}ncrotron, Caixa
Postal 6192 - CEP 13084-971, Campinas, SP , Brazil}

\date{\today}

\begin{abstract}

Understanding the electronic structure of semiconductor
nanostructures is not complete without a detailed description of
their corresponding spin-related properties. Here we explore the
response of the shell structure of InAs self-assembled quantum
dots to magnetic fields oriented in several directions, allowing
the mapping of the g-tensor modulus for the s and p shells. We
found that the g-tensors for the s and p shells show a very
different behavior. The s-state in being more localized allows the
probing of the confining potential details by sweeping the
magnetic field orientation from the growth direction towards the
in-plane direction. As for the p-state, we found that the g-tensor
modulus is closer to that of the surrounding GaAs, consistent with
a larger delocalization. These results reveal further details of
the confining potentials of self-assembled quantum dots that have
not yet been probed, in addition to the assessment of the
g-tensor, which is of fundamental importance for the
implementation of spin related applications. \end{abstract}


\maketitle

Electronic magnetism in semiconductor nanostructures is one of the
important properties to be harnessed in spintronic devices
\cite{wolf} as well as in prototypical systems for quantum
information processing\cite{divincenzo}. In order to understand
and separate the effects of quantum confinement and band
structure, including spin-orbit coupling, strain and non-parabolic
effects, the response of the electronic spin on an applied static
magnetic field can provide an improved picture of the overall
quantum system. The electronic g tensor, which describes the
symmetries and magnetic response of the unpaired electron system,
is thus a very important tool to assess and investigate these
fundamental aspects of spin electronics in nanostructures.

For conduction electrons in bulk semiconductor crystals, the
g-factor can be determined accurately by second-order
$\mathbf{k\cdot p}$ theory using Roth's
equation\cite{lax59,elliot54}, and confirmed by experiment
\cite{weisbuch77}. For unpaired electrons bound to donors,
g-tensor differences from the free atom value of 2 for the ground
state will reveal the dependence on the crystal field and
spin-orbit coupling\cite{abragam}. The symmetries of defects and
chemical environment can be also revealed by mapping the
g-tensor\cite{anderson}. In addition to that, the anisotropic part
of {\it g} influences spin-lattice relaxation and is important for
spin-related applications \cite{roth60,calero05}. For the case of
quantum wells\cite{malinowski00} and wires \cite{oestreich} the
g-tensor will be affected by quantum confinement, strain and
composition fluctuations.




Experimental investigations of g factors and g tensors have been
reported for metallic nanoparticles\cite{petta} and
lithographically defined quantum dots (QDs)
\cite{ensslin02,hanson03}. For metallic nanoparticles, the
difficulties arise in finding the symmetry axis, which can be
determined from the g-tensor mapping. In addition, because the
electron mean free path is smaller than the particle size, angular
momentum may not be a good quantum number. For lithographically
defined quantum dots, the Zeeman splitting and the orbital
splitting have comparable energy scales, thus preventing the
evaluation of the out-of-plane g-tensor component. For
self-assembled quantum dots, a number of experiments have
demonstrated striking similarities with the atomic behavior, such
as Hund's rules and the Aufbau principle in determining the shell
filling for electrons \cite{warburton98,zunger05}. The charging,
Zeeman splitting and single-particle energies are all different
for this case, which allows them to distinguished for excitons and
electrons \cite{bayer99, goni, apl02,apa03}. More recently,
calculations were carried out displaying the relationship between
the g tensor and the electronic structure for quantum
dots\cite{pryor}.

In this Letter we explore the shell structure dependent spin
properties of electrons trapped in InAs quantum dots (QDs). By
evaluating the electron addition energies inferred from
magneto-capacitance data, we present an experimental account on
the g-tensor modulus for the s and p states which were mapped out
according to the crystallographic directions of highest symmetry.


InAs QDs were grown by molecular beam epitaxy and capped with thin
InGaAs strain reducing layers, as described elsewhere \cite{apl02,
apa03}. These structures were embedded in capacitance structures
that were subsequently defined by conventional photolithography.
The area of the devices was $4\times10^{-2}\rm{mm}^2$, hence
encompassing an ensemble of about $10^8$ QDs per device.
Magnetocapacitance experiments were carried out at 2.7K for
magnetic field intensities ranging from 0 to 15T. Field sweeps
were performed at 15$^\circ$ intervals covering at least
180$^\circ$ by tilting the sample with a goniometer.

Figure 1 (a) shows the second derivative of magnetocapacitance
spectra taken for field sweeps along the [001] and [110]
directions (polar scan). The energy scale derived from the applied
bias and voltage-dependent lever arm is translated into the
chemical potential within the QDs referenced to the GaAs
conduction-band edge. The lever arm was calculated taking into
account depletion effects in the back contact\cite{apa03}, thus
allowing the determination of a precise energy scale. The gray
scale map is a representation of the density of states (DOS),
where the 0D (up to $\sim$ -80meV) and 2D (above $\sim$ -80meV)
levels associated with the QDs and wetting layer can be easily
identified. For in-plane field all orbital effects are minimized
given the pancake geometry of the quantum dots; also negligible
are the effects of the magnetic field on the wetting layer, which
exhibits Landau-level fillings for magnetic fields perpendicular
to the sample surface.

\begin{figure}[ht]
\centerline{\epsffile{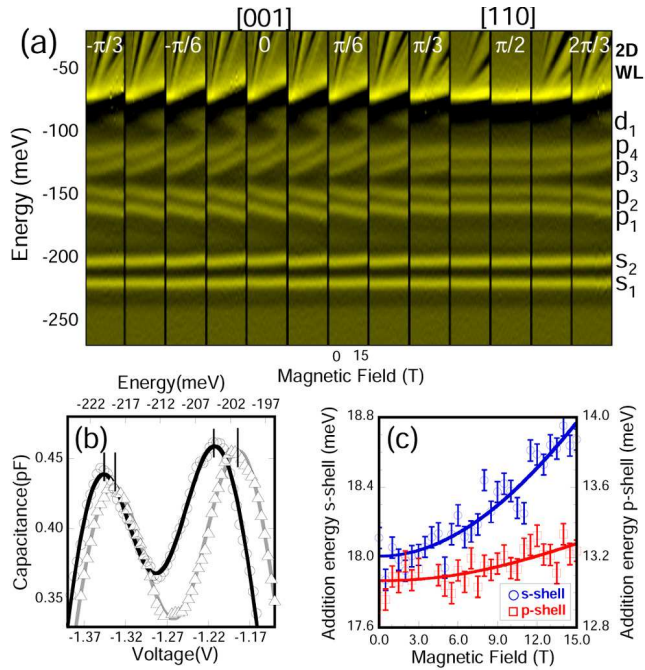}} \caption{{\bf(a)}
Measured data (derivative) for polar scan, {\it i.e.}, magnetic
field sweeping from parallel to the $[001]$ direction towards the
$[110]$ direction. The grayscale is keyed to the second derivative
of the capacitance spectra, with light colors indicating a higher
density of states. Each of the 13 frames corresponds to B sweeps
from 0 to 15T, taken at $15^{\circ}$ intervals with $\theta$ the
angle between B and [001]; The s, p and d states, as well as the
wetting layer 2D levels are indicated as well; {\bf(b)} Different
capacitance spectra showing the filling of the s shell with 1 and 2
electrons, dots correspond to $\mathbf{B}=0$T and triangle to
$|\mathbf{B}|=15$T, and $\theta=\pi/12$. The solid lines correspond
to two gaussian fits. {\bf(c)} Fits to equation \ref{eqn_bri} for s
and p shell electron addition energies for $\theta=\pi/2$.}
\label{emap}
\end{figure}


From the electron addition energies \cite{apl02,apa03,zunger05},
orbital, electrostatic and Zeeman contributions can be separated.
Figure 1 (b) shows the shift of the capacitance peaks for the
sequential charging of the s-level on the applied magnetic field
for $\theta=\pi/12$. The effects of Coulomb charging, diamagnetic
shifts and Zeeman splitting can be seen in these data. One can
separate the orbital effect contribution from the others by
subtracting the peak positions, as the diamagnetic contribution is
the same for both. For the s shell, using a Fock-Darwin (FD)
formalism\cite{warburton98,apa03} one has for the loading of the
first and second electron for T=0 and $\mathbf{B}\parallel[001]$:
$E_{s1}=E_z+\hbar\Omega-|g_{zz}|\beta\mathbf{B}/2;
E_{s2}=E_{s1}+|g_{zz}|\beta\mathbf{B}+E_{C}^s$, where $E_z$ is the
confinement along the growth direction,
$\Omega=\sqrt{\omega_{0}^2+\omega_{c}^2/4}$, with $\omega_{c}$ as
the cyclotron frequency, $E_{C}^s$ the Coulomb charging energy at
zero magnetic field for the s shell electrons, and
$|g_{zz}|\beta\mathbf{B}$ the Zeeman splitting. For the s shell,
one finds $\hbar\omega_0=37.8\pm$0.2meV and $E_z\sim-280meV$.
Detailed modeling can provide an even better description
\cite{zunger05}, however it is not necessary to capture the
essential features. Electrostatic effects can be calculated given
the single particle energies within the FD
framework\cite{warburton98}. For the s and p shells we find
$E_{ss}$=17.2meV and $E_{pp}$=13.4meV. Figure 1 (c) shows the
dependence of the addition energies for the s-s and p-p
configurations on the applied magnetic field for $\theta=\pi/2$.
We find that an agreement within 5\%-15\% can be found from the
calculated and measured addition spectra.

Two effects must be considered when analyzing these data - i)
wavefunction compression and its effect on the charging energies
and ii) temperature. As the magnetic field is raised, the
wavefunction is compressed which increases the Coulomb charging
energies. Under the FD formalism, this effect can be calculated as
$E_{C}^i(B)=
E_{C}^i(0)\left(1+\omega_c^2/4\omega_{0}^2\right)^{1/4}$, where
$E_{C}^i(0)$ is the Coulomb charging energy at zero magnetic field
for the $i$-shell. This effect takes place for all directions of
the applied magnetic field, but to a lesser extent for the
in-plane configuration due to a stronger confinement along the
growth direction. As a zeroth-order approximation, we assume this
effect to be the same for all configurations. The consequence of
this assumption is to underestimate the Zeeman contribution for
in-plane magnetic fields.


As stated above, the experiments were carried out with QDs
ensembles. Thus, an accurate description of this system at finite
temperatures requires usage of a magnetization model for a system
of $n$ non-interacting spins, which can be carried out by
calculating the partition function for the system. For the current
analysis, we take into account the temperature-dependent spin
contribution on the addition energy spectra. This yields a direct
relationship between the addition energy $\Delta \mu$, the
$B$-dependent Coulomb charging, the Zeeman splitting and the
temperature:

\begin{equation}
\Delta \mu_s  = E_{CB}(B)+
2k_BT\ln\left[2\cosh\left(\frac{g\mu_B|\mathbf{B}|}{2k_BT}\right)\right]
\label{eqn_bri}
\end{equation}

Figure 1 (c) shows $\Delta \mu$ as a function of the magnetic
field and the corresponding fit to equation \ref{eqn_bri} for the
s shell at $\theta=\pi/2$ (\emph{i.e.}, in-plane magnetic field).
For the p shell, the same description applies, and a similar
relation can be derived. However, the observed addition energies
and the dependence on $\mathbf{B}$ are smaller (fig. 1c), and it
becomes difficult to implement a reliable fit because of two
additional factors: i) a smaller Coulomb charging energy, and ii)
broadening of the capacitance peaks due to the Coulomb disorder.
Coulomb charging is given by $E_{C}^p=3/4E_{C}^s$ while Coulomb
disorder scales as $n^2$, where $n$ is the number of electrons
trapped inside the QDs \cite{gilberto97}. In order to fit the data
we assume $E_{C}^p(B)=E_{C}^p(0)$ and retained the temperature
dependence (from Eq. \ref{eqn_bri}). Although an approximation, it
is instructive to present this analysis as it represents an upper
bound on the g-tensor, and most importantly, it can help in
elucidating the symmetries for this particular state. This
reasonably simple model which takes into account the most
important factors in determining the Zeeman splitting in quantum
dots, including temperature effects and wavefunction compression,
describes well the whole data set.


From the g factor obtained at each angle and for each shell, the g
tensor modulus was determined for the polar and azimuthal scans by
$
|g_{pol}|=\sqrt{g_{[001]}^2\cos(\theta)^2+g_{[110]}^2\sin(\theta)^2}$
and $
|g_{az}|=\sqrt{g_{[110]}^2\cos(\phi)^2+g_{[1\overline{1}0]}^2\sin(\phi)^2}$.
Figures 2 and 3 show the polar and azimuthal scans. On the top
panel the experimental set-up is represented showing the
orientation of the magnetic field with respect to the QD
crystalline axis, as well as the FD wavefunctions for each shell
calculated for $\mathbf{B}=15\rm{T}$. The data are shown on the
bottom panel with the corresponding fits.

\begin{figure}[ht]
\centerline{\epsffile{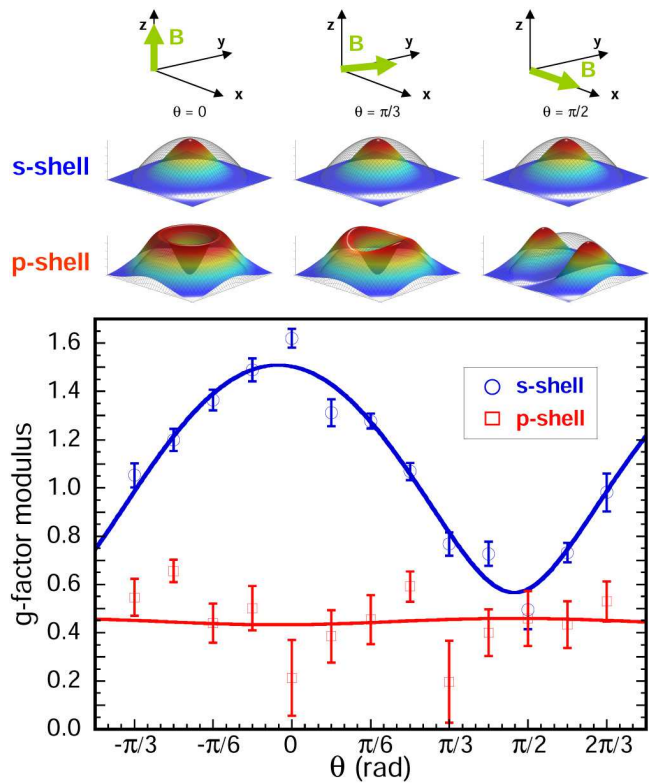}} \caption{ (top)
Wavefunction for the s and p shell for three directions on applied
magnetic field $\theta = 0$, $\theta = \pi/3$ and $\theta = \pi/2$.
$x$,$y$ and $z$ correspond to $[110]$, $[1\overline{1}0]$ and
$[001]$ directions; a QD is schematically represented in the same
plot. (bottom) g-tensor for the s and p shells for the polar scan.}
\label{polar_fig}
\end{figure}

\begin{figure}[ht]
\centerline{\epsffile{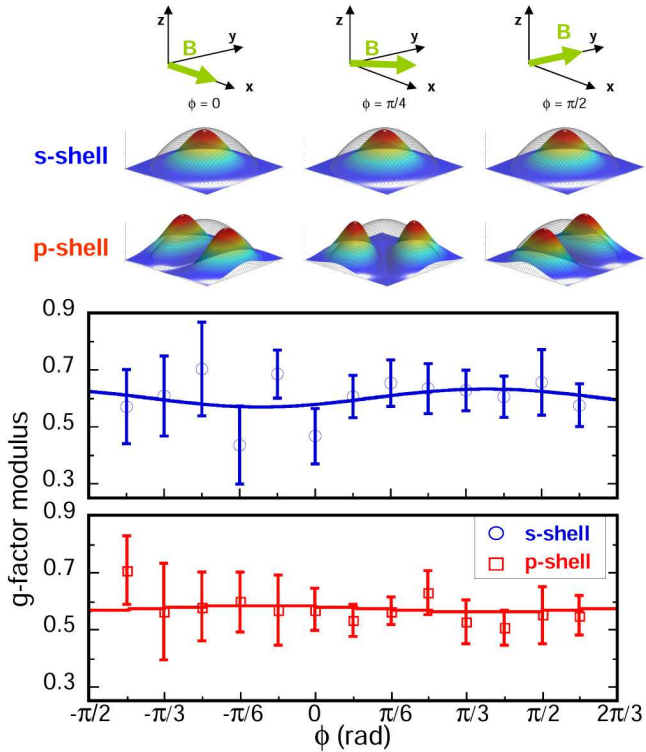}} \caption{(top)
Wavefunction for the s and p shell for three directions on applied
magnetic field $\phi = 0$, $\phi = \pi/4$ and $\phi = \pi/2$.
$x$,$y$ and $z$ correspond to $[110]$, $[1\overline{1}0]$ and
$[001]$ directions; a QD is schematically represented in the same
plot. (bottom) g-tensor for the s and p shells for the azimuthal
scan.} \label{azimuthal_fig} \end{figure}

From the polar scan one can immediately note that for the s shell
the g-factor is quite anisotropic whereas for the p shell it is
constant and always smaller. This is a quite surprising result at
first, considering the the single particle energy
($\hbar\omega_0=37.8$meV) and confinement along z ($E_z=-280$meV),
and the fact that the z-component of the wavefunction for both
states is basically the same. From the fit, we find
$|g_{[001]}^s|=1.51\pm0.03$ and $|g_{[110]}^s|=0.57\pm0.05$, and
$|g_{[001]}^p|\approx |g_{[110]}^p|=0.47\pm0.07$. The larger
values for $g^s$ are consistent with a stronger localization
inside the QD due to the deeper confinement. In order to draw a
qualitative picture for the wavefunctions, we plot the solutions
for the s and p wavefunctions for different magnetic field tilt
angles (figures 2 and 3, top panel). Insofar as the behavior for
$|g^s|$ is concerned, one notes that the wavefunction is more
localized into the QD and consequently more sensitive to the
confinement potential details at the QD center. It is expected a
high anisotropy given the pancake geometry of the QDs. The
somewhat unexpected result comes about for the p shell. A highly
isotropic behavior is found, and we associate this behavior to the
symmetry of the p- wavefunction, as depicted in the upper panel of
figure 2. First, as the electrons on the p shell are more
delocalized, a leakage of the wavefunction along the growth
direction takes place, which brings the g-factor modulus values
closer to that of the matrix ($|g_{GaAs}=0.44|$). Second, the
wavefunction for the p shell has a node at the center of the QDs,
which permits probing of the regions outside or at the interfaces
of the QDs. Hence, by measuring the g-factor for the s and p
shell, one can evaluate the details of the confining potential at
selected spatial locations, in a similar fashion that was carried
out for the determination of the chemical environment of deep
levels \cite{anderson}.

 Figure 3 shows the g-tensor for an in-plane
field configuration (azimuthal scan). For both s and p shells the
g-factor is independent of the field direction within the
experimental uncertainties, consistent with a cylindrical symmetry
for the QDs.

Several competing effects have to be taken into account when
interpreting the obtained results: i) local strain and local
crystal fields which change the details of the confining
potential, ii) non-uniform composition, iii) quantum confinement,
and iv) non-parabolicity, which takes place in InAs. All these
parameters may influence the spin-orbit coupling, which is one of
the important components in the g-factor determination.
Modifications into g-factor has been demonstrated by tuning both
strain \cite{apa03,nakaoka05} and composition \cite{Bjork05} in
the QDs. If one compares these experimental results with theory, a
good agreement is found which corroborates the description by
Pryor {\it et al} \cite{pryor}. In essence, we find that effects
i-iii are basically the same, and they by and large determine the
behavior for the g-tensor. More careful experiments on different
samples are required to draw a more complete picture on the
relative effect of each component.

\begin{figure}[th]
\centerline{\epsffile{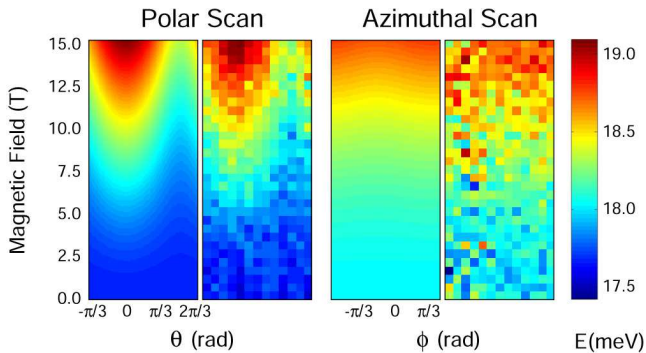}} \caption{(left)
calculated and measured addition energies for the polar scan;(right)
same for the azimuthal scan.} \label{map}
\end{figure}

As a final verification of the models utilized in this work, we
calculate the electron addition energy for n=2, {\it i.e.,}
s shell filling, from the values obtained from the fits. The
addition energy can be calculated according to equation
\ref{eqn_bri}. Figure \ref{map} shows the calculated (left) and
measured (right) results for the polar and azimuthal scans. The
usefulness of this verification is primarily for evaluation of
issues such as adequacy of the proposed model, signal to noise
ratio, and an overall picture of what can be resolved for this two
electron system. An important point mentioned previously was that
of ensemble measurements. In this experiment, one obtains a more
reliable evaluation of the g-tensor as one is averaging over many
different spin configurations which are temperature and
magnetic-field dependent. This experiment thus provides a more
representative description of g-tensor in nanostructures.

In summary, we have inferred the g-tensor for the s and p shells
of self-assembled QDs. We found that for the s shell the g-tensor
is highly anisotropic, reflecting the confinement potential
details. For the in-plane component and for the p shell, the
modulus of the inferred g factors were close to the bulk GaAs
value. Finally, we found that for the p shell the g tensor was
isotropic within our experimental resolution, which is consistent
with wavefunction having a node at the QD center and being more
delocalized along the growth direction. We acknowledge the
financial support by CNPq, FAPESP and HP Brazil, and the usage of
the high magnetic field facility at GPO/IFGW - UNICAMP.



%

\end{document}